\documentclass[useAMS,usenatbib]{mn2e}
\usepackage{graphicx}
\usepackage{epstopdf}

\def\Mpc{\hbox{$\rm\thinspace Mpc$}}

\def\hMpc{\hbox{$\thinspace h^{-1}\Mpc$}}

\def\msun{\hbox{${\rm\thinspace M}_{\sun}$}}

\title[Redshift-space limits of bound structures]{Redshift-space limits of bound structures}
\author[R. D\"unner et al.]{Rolando D\"unner$^1$,  Andreas Reisenegger$^{1,2}$, Andr\'es Meza$^{3,4}$\thanks{Researcher of the Academia Chilena de Ciencias 2004-2006}, Pablo A. Araya$^5$,
\and Hern\'an Quintana$^1$\\
$^1$Departamento de Astronom\'ia y Astrof\'isica, Facultad de F\'isica, Pontificia Universidad Cat\'olica de Chile, Casilla 306, Santiago 22, Chile\\
$^2$European Southern Observatory, Alonso de C\'ordova 3107, Vitacura, Santiago, Chile\\
$^3$Departamento de F\'isica, Facultad de Ciencias F\'isicas y Matem\'aticas,  Universidad de Chile, Casilla 487-3, Santiago, Chile.\\
$^4$Departamento de Ciencias F\'{\i}sicas, Universidad Andr\'es Bello, Av. Rep\'ublica 252, Santiago, Chile\\
$^5$Kapteyn Astronomical Institute, University of Groningen, P.O. Box 800, 9700 AV Groningen, The Netherlands}

\begin{document}

\date{November 2006}

\pagerange{\pageref{firstpage}--\pageref{lastpage}} \pubyear{2006}

\maketitle

\label{firstpage}

\begin{abstract}

An exponentially expanding Universe, possibly governed by a cosmological constant, forces gravitationally bound structures to become more and more isolated, eventually becoming causally
disconnected from each other and forming so-called ``island universes". This new scenario reformulates the question about which will be the largest structures that will remain gravitationally bound, together with requiring a systematic tool that can be used to recognize the limits and mass of these structures from observational data, namely redshift surveys of galaxies. Here we present a method, based on the spherical collapse model and $N$-body simulations, by which we can estimate the limits of bound structures as observed in redshift space. The method is based on a theoretical criterion presented in a previous paper that determines the mean density contrast that a spherical shell must have in order to be marginally bound to the massive structure within it. Understanding the kinematics of the system, we translated the real-space limiting conditions of this ``critical" shell to redshift space, producing a projected velocity envelope that only depends on the density profile of the structure. From it we created a redshift-space version of the density contrast that we called ``density estimator", which can be calibrated from $N$-body simulations for a reasonable projected velocity envelope template, and used to estimate the limits and mass of a structure only from its redshift-space coordinates.

\end{abstract}


\section{Introduction}\label{intro}
The current paradigm of an exponentially expanding Universe
implies that large-scale, gravitationally bound structures will
eventually become causally disconnected from each other, forming
island universes scattered inside a mostly empty Universe (e.~g.,
\citealt{Adams,Tzihong,Nagamine2003,Busha}). In our previous paper
(\citealt{Dunner1}, hereafter ``Paper I"), we presented a
criterion to determine the limits of bound structures, defining
superclusters as the biggest gravitationally bound structures that
will be able to form. This criterion defined a critical density
contrast over which a spherical shell will stay bound to a
spherically distributed overdensity. As defined, this criterion
can only be applied to data given in real, three-dimensional space
(``real space"). This is not the case of observational data, which
comes from large galaxy surveys having two angular coordinates and
a velocity coordinate. Using the Hubble law, one can estimate the
real distance to an object from its recession velocity, but, given
that we are interested in dense, relatively evolved structures
with significant peculiar velocities, our estimation will be
strongly affected by the velocity dispersion of the structure,
fooling any attempt to apply a real-space-based method
\citep{Kaiser87}.

Here we present a way to apply our theoretical criterion to
redshift-space data, permitting its application to redshift
surveys. The new criterion is based on the geometrical appearance
of the real-space criterion as seen in redshift space, and needs
to be calibrated using statistics from $N$-body simulations to
account for velocity dispersions not considered in the theoretical
model presented in Paper I. This method represents an
alternative to the caustic approach, first proposed by
\cite{Regos89} and further developed by \cite{Diaferio97} and
\cite{Diaferio99}. The caustic method, based on the direct search
for caustic curves which represent the redshift-space envelope of
the bound structure within it, has been extensively used to study
galaxy clusters (e.~g.,
\citealt{Haarlem93,Geller99,Maze,Rines00,Rines02,Rines03,Biviano03,Diaferio05,Rines06}),
constituting the most widely used method in the area. Among its
main achievements, it has been used to measure the mass profile
and light to mass radio from galaxy clusters.

Our method, even though it shares some of the basic
elements of the caustic method, as will be discussed later, has
the independent motivation of directly representing the spherical
collapse density criterion for the critical shell in redshift
space, giving a clear physical interpretation to its results.

In Section \ref{ingredients}, we discuss the effects of
transforming the real-space data into redshift space as seen from
$N$-body simulations, introducing the critical projected velocity
envelope, which is a theoretical construction produced by joint
projection of all shells, within a certain radius, intersecting
the line of sight. We also study the implementation of a
parameterized template for the density profile, for which we used
the density profile proposed by \cite{NFW}.

In section \ref{method}, we propose a new method for applying the
criterion in redshift space, which is presented in three
alternative versions. For this we introduce the concept of
``density estimator", which replaces the previously used density
contrast in determining the threshold that defines the critical
shell. We statistically determine the value of this estimator,
together with analyzing the main divergences from the spherical
collapse theoretical model that introduce errors into the method.
Each version of the method will require one or more density
estimators, which will be presented later.

In section \ref{testing}, we test our method using our $N$-body
simulation, estimating the systematic error that is expected for
radius and mass estimations for the gravitationally bound
structure.

Finally, in section \ref{Conclusiones}, we present our
conclusions, together with a step by step recipe for applying one
of the proposed methods.

\section{Ingredients for a Fitting Method}\label{ingredients}
\subsection{Redshift-Space Representation of the Spherical Collapse Model}\label{theory}

In Paper I, we showed that the spherical collapse model, when
extended to the case of a universe dominated by a cosmological
constant, can be used to set a criterion for the ``critical"
(marginally bound) shell of a mass concentration. The spherical
shells are characterized by a single parameter named density
contrast $\Omega_s$, where the $s$ stands for shell. Its value,
for the critical shell (cs), can be written as
\begin{equation}
\Omega_{\mathrm{cs}} = \frac{\bar{\rho}^{\mathrm{cs}}_\mathrm{m}}{\rho_\mathrm{c}} = 2.36,
\label{real-space-crit}
\end{equation}
where $\bar{\rho}_\mathrm{m}$ is the mean mass density enclosed by the
shell, and $\rho_\mathrm{c}=3H^2_0/8\pi G$ is the critical density of the
Universe.

In the simulations, this criterion was shown to give an external
limit to the extension of gravitationally bound structures,
overestimating their mass by 39\% on average. Nonetheless, the
model gives in-falling velocity predictions which correctly follow
the lower envelope of radial velocities deep into the virialized
core of bound structures (see Fig. 8 in Paper I). This, together
with the one-to-one relation between the shell's infall speed and
its enclosed density contrast, makes it possible to extend the
model to a redshift-space representation.

\begin{figure}
\centering
    \centering
    \includegraphics[width = 41mm]{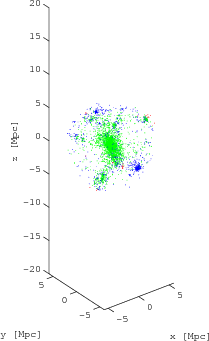}
    \includegraphics[width = 41mm]{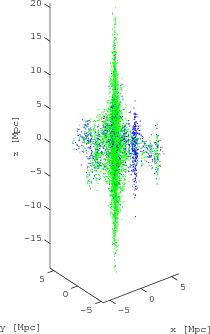}
    \caption{Three-dimensional effect of transforming to redshift space using
the Hubble equivalence between velocity and distance. On the left,
structure in real space. On the right, structure in redshift
space. In green particles bound to the structure and in blue
unbound particles. For clarity, only particles inside the critical
radius in real space were plotted.}
    \label{fig:real-redshift}
\end{figure}

\begin{figure}
\centering
    \centering
    \includegraphics[width = 60mm]{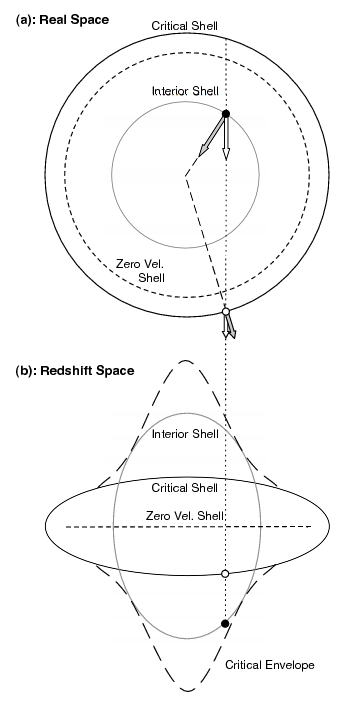}
    \caption{Redshift-space appearance of spherical shells. Panel (a) shows the position of
shells in real space: the interior shell (grey) is contracting, the dashed shell is at rest, and the outer shell (black) is expanding at the critical velocity. The black dot on the contracting shell is falling faster than the expansion of the white dot on the critical shell, so its projected velocity is higher, appearing closer in redshift space. Panel (b) shows the corresponding ellipsoids in redshift space: the colouring has been kept from panel (a). The critical shell appears shrunk along the line of sight because its expansion speed is less that the Hubble flow at its radius. The interior shell instead appears elongated along the line of sight because its contraction speed is higher than the Hubble flow at its radius, but also flipped, such that the far side appears closer that the near side. The zero velocity shell appears as an horizontal line. The black long-dashed line shows the projected velocity envelope obtained from including all shells inside the critical shell.}
    \label{fig:perfil}
\end{figure}

To go from the real-space to the redshift-space representation, we
need to replace the coordinate along the line of sight by the
corresponding recession velocity. For simplicity, we will assume
that the distance to the structures is much greater than their
size, so we do not need to account for angular effects and the
replacement can be done directly in Cartesian space. An example of
this transformation is shown in Figure \ref{fig:real-redshift},
where the structure on the right is the redshift representation of
the structure on the left, as seen by an observer looking along
the $Z$ axis. In the case of a spherically symmetric structure, as
described by the spherical collapse model, the projected velocity
seen by an observer will be composed by the projection of the
radial velocity of the spherical shell with respect to its center
and the recession velocity of the whole structure (see Figure
\ref{fig:perfil}A).

Let us consider a spherically symmetric, gravitationally bound
structure. Each shell has its own radial speed, which only depends
on its enclosed density contrast. As the spherical collapse model
constrains the shells not to cross each other, we will see the
innermost shells falling at greater speeds than outer ones. When
moving from the center to greater radii, we will cross a shell
that just stopped its expansion, through slowly expanding ones, up
to shells expanding with the Hubble flow.  According to this
model, the critical shell will be expanding with a speed (critical
velocity) of only 29\% of the Hubble flow at the present time (see
Paper I).

We are interested in identifying all shells lying within the
critical shell, each of which will have a different projected
shape in redshift space. Slowly expanding shells will appear as
ellipsoids shrunk along the line of sight, while fast contracting
ones will appear elongated along the line of sight, but flipped,
so that apparently closer points will really be at the more
distant side of the structure \citep{Kaiser87}. A diagram
explaining this effect is shown in Figure \ref{fig:perfil}, where
a fast contracting shell has a higher projected velocity than the
outer expanding shell, protruding from the outer ellipsoid and
producing the well-known ``Finger of God" effect.

We define the {\it projected velocity envelope} as the surface
that encloses the redshift-space representation of all shells
within a given radius $R$. This can be written as
\begin{equation}
\label{v_p}
v^{\mathrm{env}}_{\mathrm{p}}\left(r_\mathrm{p}\right)=\max_{r\in[r_{\mathrm{p}}, R]}\left\{|v_r\left(r\right)|\sqrt{1-\left(\frac{r_p}{r}\right)^2}\right\},
\end{equation}
where $v_\mathrm{r}$ is the radial velocity respect to the centre
of the structure and $r_\mathrm{p}$ is the projected radius on the
sky (\citealt{Maze}). The spherical collapse model allows to
predict $v_\mathrm{r}(r)$, and therefore
$v^{\mathrm{env}}_{\mathrm{p}}\left(r_\mathrm{p}\right)$, directly
from the density profile (Paper I). Setting $R$ equal to the
critical radius of the structure ($r_\mathrm{cs}$), yields the
critical projected velocity envelope (``critical envelope", for
short), corresponding to the redshift-space envelope of all shells
within the critical shell. The velocity envelope, when
defined up to the turn-around radius, is equivalent to the
caustics defined by \cite{Regos89}.

If all the assumptions of the spherical collapse model were true,
one would expect that all the bound particles should lie within
this critical envelope. Moreover, if inner ellipsoids protrude
from the ellipsoid defined by the critical shell, then one would
expect to see contamination in the protruding regions from objects
outside the critical shell.

An effect that is not considered in the spherical collapse model
are the velocity perturbations due to the interaction between
nearby objects that will mix many particles into and out of the
critical envelope, causing large systematic and random errors in
the intended criterion for determining the limits of the bound
structure in redshift space. To account for these systematic
errors we used $N$-body simulations as described in the next
section.

\subsection{Velocity Envelope in $N$-Body Simulations}\label{simulations}

In order to study the behaviour of structures in redshift space,
we used numerical simulations, which permitted us to observe the
redshift-space distribution of the structures, knowing at the same
time which particles were bound and which were not. Our
simulations are the same as described in Paper I, performed with
the GADGET2 code \citep{Springel2}, containing $128^3$ dark matter
particles inside a box of side length $100\hMpc$, and considering
a flat $\Lambda$CDM universe with $\Omega_\Lambda = 0.7$. We took
snapshots at the present time ($a=1$) and in the far future
($a=100$), assuming that in late epochs the structure evolution
will decrease significantly so no major changes will be seen from
then on (see also \citealt{Busha} and \citealt{Nagamine2003}).

We selected the 11 largest structures for our study, with masses
ranging from $1\times10^{14} \msun$ to $7\times10^{14} \msun$.
Although these structures might be rather small to represent our
current understanding of superclusters, many of them showed
significant substructure, as expected from objects that are still
evolving into a virialized state. The bound particles were
identified using the state at $a=100$ and then correspondingly
tagged and followed to the present frame, repeating the procedure
described in Paper I.

In order to transform the simulated data to redshift space, we
replaced the distance along one axis by the corresponding
projected velocity. This can be done in any direction, but for
simplicity we did so only along the three main axes, giving a
total of 33 data sets for statistical analysis.

\begin{figure}
    \includegraphics[width = 84mm, trim = 18 5 40 18]{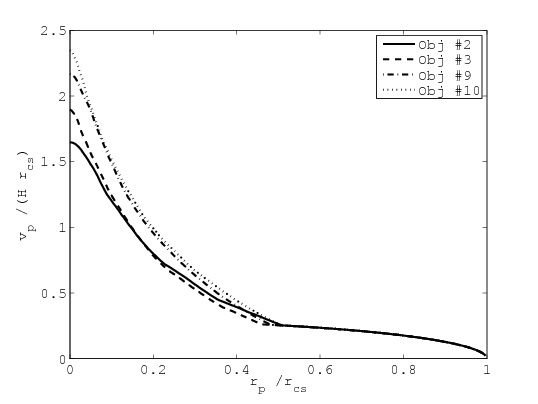}
    \caption{Critical projected velocity envelopes along the line of sight as
expected in the spherical collapse model from the density profile
for four well-behaved, representative simulated structures. Badly
behaved structures, generally implying double cores or very
significant substructure, usually present lower peak velocities
and wider central regions. Radii are normalized in terms of the
critical shell's radius and velocities in terms of the Hubble flow
at that radius. The masses satisfy $M_\mathrm{obj\#2} >
M_\mathrm{obj\#3} > M_\mathrm{obj\#9} > M_\mathrm{obj\#10}$ (see
Table \ref{NFW_M}).}
    \label{fig:Vp}
\end{figure}

Figure \ref{fig:Vp} shows the critical envelope for two objects in
our study with nearly spherical symmetry, with radii normalized in
terms of the critical radius $r_\mathrm{cs}$ and velocities
referred to the Hubble flow at $r_\mathrm{cs}$. These critical
envelopes were obtained measuring the density contrast at all
radii and applying the procedure described in Paper I to find the
radial velocity, and finally projecting it along the line of
sight. At large radius, where there is no ellipsoid crossing, we
see the ellipsoid corresponding to the critical shell, which has
the same shape since its expansion velocity is the same fraction
($29\%$ in present time) of the Hubble flow (see Paper I). At
smaller radius we observe the existence of ellipsoid crossing in
redshift space, which is produced because the inner shells are
contracting faster than the expansion velocity of the critical
shell. Adjusting the object's centre, we observe an improvement in
matching the resulting profiles. Specifically, a choice closer to
the densest core of the structure increases the height of the peak
of the velocity envelopes and improves the agreement between the
shapes of curves corresponding to different objects. Even though
the profiles do not match exactly at small radii, all of them
share the same characteristic shape, showing higher velocities in
less massive objects, meaning a higher concentration than more
massive structures, in qualitative agreement with the results of
\cite{NFW}. These properties suggest that it may be possible to
estimate the critical envelope by a general template that depends
only on the bound mass of the structure.

To study the accuracy of the prediction given by the critical
envelope for the location of particles in redshift space, we
counted the number of bound particles inside and outside the
critical envelope, as well as the number of unbound particles
inside the critical envelope. Our results are summarized in Table
\ref{performance}. We used the same statistical indicators as in
Paper I, so now we can compare the performance of the criteria in
redshift and real spaces. The indicators, all expressed as {\it
percentages of the total number of particles inside the critical
envelope}, are:
\begin{itemize}
\item{$A$: particles inside the critical envelope that do not belong to the cluster at $a = 100$.}
\item{$B$: particles inside the critical envelope that do belong to the cluster at $a = 100$.}
\item{$C$: particles outside the critical envelope that belong to the cluster at $a = 100$.}
\item{$D$: all particles that belong to the cluster at $a = 100$.}
\end{itemize}
Note that $A+B$ = 100\% and $D = B+C$.

\begin{table}
  \caption{Mean values and standard deviations for several performance
indicators for the critical envelope corresponding to the true
density profile (see text).}
  \centering
  \begin{small}
  \begin{tabular}{| c | r  r | r  r |}
  \hline
   & \multicolumn{2}{| c |}{Redshift Space} & \multicolumn{2}{| c |}{Real Space}\\
 Indicator  & Mean & Std. Dev & Mean & Std. Dev\\
  \hline
   $A$   & 23.96 &   10.01 & 28.2 & 13.0\\
   $B$   &  76.04 &   10.01 & 71.8 & 13.0\\
   $C$   &  30.59 & 13.24 & 0.26 & 0.23\\
   $D$   & 106.62 & 17.56 & 72.0 & 13.1\\
    \hline
  \end{tabular}
  \end{small}
  \label{performance}
\end{table}

We observe that, compared to the real-space criterion (Paper I),
the new redshift-space criterion produces better results for the
indicators $A$ and $B$, but the indicator $C$ increases
significantly (see Table \ref{performance}), implying that it does
not give an external limit to the location of particles in
redshift space, in contrast with the real-space case.

As pointed out before, compared to predictions done in real space,
predictions done on redshift space are affected by a higher mixing
of bound and unbound particles. This is due to the velocity
dispersion, which makes many particles (inside the critical shell
in real space) fall outside of the critical envelope in redshift
space. The velocity dispersion is produced by local interactions
between nearby objects as they decelerate driven by the central
attractor, manifesting as local peculiar velocities in random
directions. To check if the velocity dispersion was accompanied by
an increase in the mean speed of the particles (as expected), we
plotted the later with respect to their radii, comparing it to the
absolute value of the theoretical radial velocity profile.
\begin{figure}
\centering
    \centering
    \includegraphics[width = 84mm, trim = 15 5 40 18]{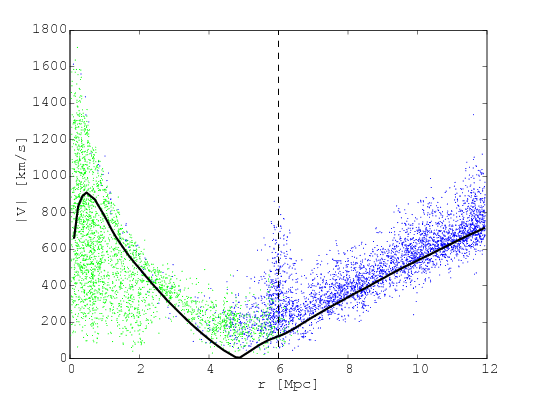}
    \caption{Absolute value of the velocities versus their radii to the centre
for Obj. \#9. The solid line shows the absolute theoretical
velocity. The dashed line indicates the critical radius. The turn-around radius is clearly observed where the solid
line touches zero, accompanied, as expected, by a minimum of
absolute velocities. For clarity we plotted a 70\% random sample
of all particles.}
    \label{fig:V_vs_r}
\end{figure}
In Figure \ref{fig:V_vs_r}, we can see how the theoretical profile
approximately marks the lower bound of absolute velocities for
particles inside the critical radius, but outside the virialized
core (around $2.5\Mpc$ from the centre), so we can claim that the
main contributor for particles escaping from the critical envelope
is the overall increment in particle speeds due to interactions
between them.

\subsection{Density Profile Template} \label{Fitting the Density Profile}

Until now, we have used the measured density profile to estimate
the critical envelope from our simulated structures. When
confronted with observational data in redshift space, we will not
have this information, so we need a general way to estimate the
density profile of a structure based on a small number of
parameters. We have seen in \S \ref{simulations} that the critical
envelopes for different structures have a coherent shape (see Fig.
\ref{fig:Vp}), but with scales correlated with the bound mass of
the structure. This is in agreement to what was proposed by
\cite{NFW}, who presented the now well-known NFW profile, which is
a generalized density profile for virialized clusters, obtained
empirically from many simulations. In particular, we used the
results from a later publication (\citealt{ENS}), which corrected
several details from the original work. In general words, they
postulate that a cluster's density profile can be estimated using
as a single parameter, the radius $r_\Delta$ at which a
characteristic (``virial") density contrast $\Delta$ is reached.
The parameter $\Delta$ depends on the cosmology and can be
obtained in a flat universe as $\Delta =
178\Omega_\mathrm{m}^{0.45}$, which in our case gives $\Delta =
103.5$. This characteristic radius is much smaller than our
critical radius, falling near the edge of the virialized zone
where clusters are less affected by external structures and look
more homogeneous, making it easier to estimate using redshift
data. The NFW profile was formulated to give good results for
radii ranging from $0.1r_s$ to $10r_s$, where $r_s =
r_\Delta/\mathrm{c}_\Delta$ and $c_\Delta$ is the concentration
parameter, which can be obtained as a function of the mass
contained within $r_\Delta$, and whose value is around $\sim 8$
for the masses of the structures studied by us. Considering that
the critical radius is of order $3-5r_\Delta$, the NFW profile is
more accurate for the inner part of the structure. This region is
the most interesting one for our purposes, since it is where
ellipsoid crossing takes place and the shape of the critical
envelope is directly dependent on the density profile. On the
other hand, we caution that large bound structures may generally
not be dominated by clusters as relaxed and spherical as those for
which the NFW profile was derived.

The NFW density profile was fitted to every object in our
simulation using the actual density in real space.

\begin{figure}
\centering
    \centering
    \includegraphics[width = 88mm, trim = 20 0 30 0]{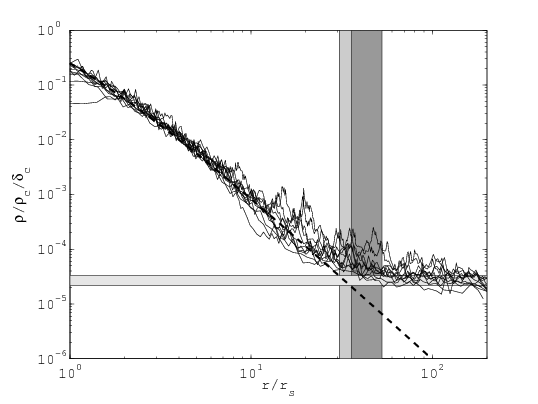}
    \caption{Density profiles of all structures (solid), compared to the NFW
template (dashed). The densities and radii have been
normalized by their corresponding scale in the NFW template to put
them on the same plot. The dark and light grey vertical areas denote placement
of critical radii measured from the simulated data and from the
fitted NFW template respectively (the dark grey area actually extends
under the light grey area up to its left end, but is hard to see in the
plot). The horizontal area denotes the range for the asymptotic values for
the density, $\Omega_\mathrm{m}\rho_\mathrm{c}$.}
    \label{fig:NFW}
\end{figure}
Figure \ref{fig:NFW} shows the density profiles for all
our simulated structures, compared to the NFW template profile.
Scales have been normalized by their corresponding NFW scale, in
order to let us compare all profiles to a single NFW template.
The dark and light grey vertical areas show
where the critical radii obtained from the simulated data and from
the NFW profile, respectively, are located.
The horizontal area shows the asymptotic values expected as the density
reaches the mean density of the Universe. 
Clearly, the true density profiles depart from the template at
radii somewhat smaller than the critical radius, as they approach
the mean density of the Universe. This early departure from the
model implies that estimation of the critical radius done directly
from the NFW template will be biased to lower values. Below, we
will present ways to deal with this bias.

\begin{figure*}
\centering
    \centering
    \includegraphics[width = 88mm, trim = 20 0 30 0]{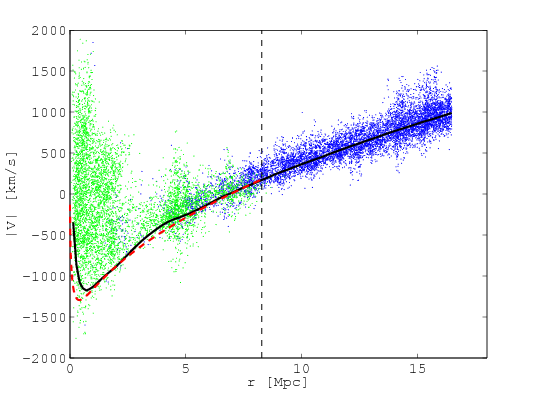}
    \includegraphics[width = 88mm, trim = 20 0 30 0]{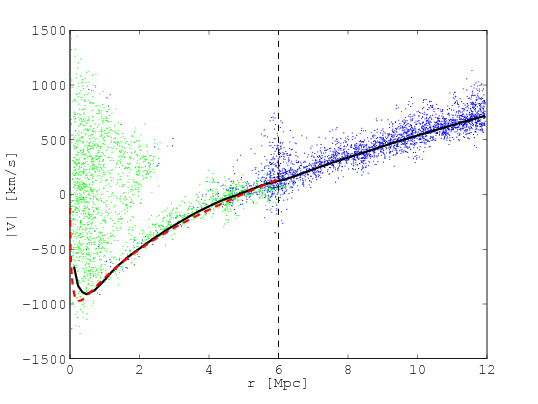}
    \caption{Radial velocities compared to the theoretical velocity profile
obtained from the true density profile (black solid line) and from
the NFW density profile fitted to the structure (red dashed line).
Green dots indicate bound objects and blue dots the opposite. On
the left object \#1, and on the right object \#9. Sampling 50\%.}
    \label{fig:Vr_NFW}
\end{figure*}
\begin{figure*}
    \includegraphics[width = 88mm, trim = 20 0 30 0]{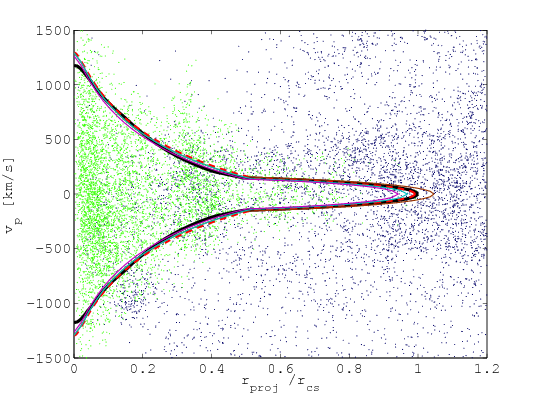}
    \includegraphics[width = 88mm, trim = 20 0 30 0]{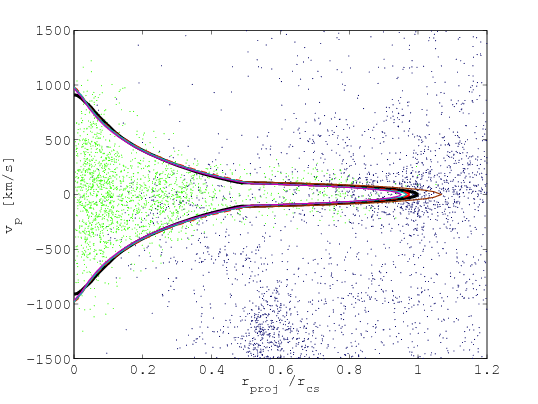}
    \caption{Projected velocities versus normalized projected radius. In green, particles that will eventually fall in the object. In blue, particles that will escape. The curves show different critical envelope estimations: The black solid line shows the true velocity envelope. The red dashed line is the NFW envelope adjusted using the observed $r_\Delta$ value. The cyan solid line is the NFW envelope fitted using the NFW-core method (fitting methods are explained in \S \ref{recipes}). The magenta solid line is the NFW envelope fitted using the NFW-cs method. The brown solid line show the combined envelope fitted using the combined method. On the left object \#1, and on the right object \#9. Sampling 50\%.}
    \label{fig:perfiles9}
\end{figure*}
As seen in Figures \ref{fig:Vr_NFW} and \ref{fig:perfiles9}, the
radial velocities and critical envelopes predicted using a NFW
density profile show different levels of agreement with the true
velocities depending on the range of radii considered. For small
radii, it yields a higher density than observed, predicting higher
infall velocities. Given the resolution of our simulation, which
was intended to search for large scale structure rather than
replicating the behaviour of the virialized cores of structures,
we believe that our simulated data is unable to produce accurate
densities at such radii. For intermediate radii, the prediction is
very good, as expected, since the NFW profile was fitted at a
measured $r_\Delta$ in this range. Finally, for larger radii,
closer to $r_\mathrm{cs}$, the generalized profile gives better or
worse results depending on the absence or presence of significant
substructure, respectively. In general, the presence of
substructure increases the density at higher radii, so the NFW
profile underestimates it, consequently giving a low estimate for
the critical radius.

\begin{table}
  \caption{Critical radii considering different models in real space.
Columns: (1) Object index. (2) $r_\Delta$: virialization radius
needed to fit the NFW density profile. (3) $r_\mathrm{cs}$:
critical radius applying to our criterion ($\Omega_\mathrm{cs} =
2.36$) to the true density profile. (4)
$r_\mathrm{cs}^\mathrm{NFW}$: critical radius applying our
criterion to the fitted NFW profile
($\Omega_\mathrm{cs}^\mathrm{NFW} = 2.36$). (5) Ratio between
$r_\mathrm{cs}^\mathrm{NFW}$ and $r_\mathrm{cs}$.}
  \centering
  \begin{small}
  \begin{tabular}{|  r  c  c  c  c  |}
  \hline
  Obj.\# &  $r_\Delta [Mpc]$  &  $r_\mathrm{cs} [Mpc]$ & $r_\mathrm{cs}^\mathrm{NFW} [Mpc]$ & $r_\mathrm{cs}^\mathrm{NFW}/r_\mathrm{cs}$\\
  \hline
1   &  1.81  &   8.28 &   8.21  &  0.99\\
2   &  1.51 &   7.84 &   6.79  &  0.87\\
3   &  1.48  &   7.54 &   6.65  &  0.88\\
4   &  1.56  &   7.92 &   7.02  &  0.89\\
5   &  1.30  &   7.38 &   5.80  &  0.79\\
6   &  1.21  &   7.93 &   5.38  &  0.68\\
7   &  1.48  &   7.63 &   6.64  &  0.87\\
8   &  1.30  &   6.29 &   5.80  &  0.92\\
9   &  1.31  &   6.01 &   5.85  &  0.97\\
10 &  1.27  &   5.67 &   5.66  &  1.00\\
11 &  1.23  &   5.37 &   5.49  &  1.02\\
    \hline
  \end{tabular}
  \end{small}
  \label{NFW_R}
\end{table}

\begin{table}
  \caption{Enclosed masses considering different models in real space,
contrasted to the effectively bound mass. Columns: (1) Object
index. (2) $M_\mathrm{bound}$: Bound mass at a = 100. (3)
$M_\mathrm{cs}$: Mass enclosed by the critical radius
$r_\mathrm{cs}$. (4) Mass ratio between bounded mass and
$M_\mathrm{cs}$ (5) $M_\mathrm{cs}^\mathrm{NFW}$: Mass enclosed by
the critical radius obtained using the NFW density profile
$r_\mathrm{cs}^\mathrm{NFW}$. (6) Mass ratio for
$M_\mathrm{cs}^\mathrm{NFW}$.}
  \centering
  \begin{small}
  \begin{tabular}{| r  c  c  c  c  c |}
  \hline
  Obj.\# & $M_\mathrm{bound}$  & $M_\mathrm{cs}$ & $\frac{M_\mathrm{bound}}{M_\mathrm{cs}}$ &  $M_\mathrm{cs}^\mathrm{NFW}$ & $\frac{M_\mathrm{bound}}{M_\mathrm{cs}^\mathrm{NFW}}$\\
  \hline
1   & 6.34  &  7.64  &  0.83  &  7.60  &  0.83\\
2   & 5.29  &  6.47  &  0.82  &  5.96  &  0.89\\
3   & 4.71  &  5.76  &  0.82  &  5.26  &  0.90\\
4   & 4.14  &  6.68  &  0.62  &  5.99  &  0.69\\
5   & 4.07  &  5.41  &  0.75  &  4.31  &  0.94\\
6   & 3.23  &  6.70  &  0.48  &  3.01  &  1.07\\
7   & 2.95  &  5.98  &  0.49  &  4.53  &  0.65\\
8   & 2.44  &  3.34  &  0.73  &  3.18  &  0.77\\
9   & 2.40  &  2.91  &  0.83  &  2.79  &  0.86\\
10 & 2.02  &  2.45  &  0.82  &  2.44  &  0.82\\
11 & 1.46  &  2.08  &  0.70  &  2.18  &  0.67\\
    \hline
  \multicolumn{6}{| l |}{All masses in units of $10^{14}\msun$}\\
  \hline
  \end{tabular}
  \end{small}
  \label{NFW_M}
\end{table}

Tables \ref{NFW_R} and \ref{NFW_M} compare the predicted critical
radius and enclosed mass for every object in the analysis. We
observe that the critical radius according to the NFW profile is
89.8\% of the measured critical radius. Thus, considering that the
spherical collapse criterion gives an external limit for the
critical radius, estimations done with the NFW profile will
underestimate the size of the structure as defined by the
spherical-collapse criterion.

Concerning the mass of the structures, we find that the mean true
bound mass ($M_\mathrm{bound}$) is 82.7\% of the mass enclosed by
the NFW profile critical radius in real space. This should be
compared to the same relation for the true critical radius, where
the true bound mass is 71.7\% of the mass enclosed by it (see
Paper I for details).

\section{Fitting Method Definition}\label{method}

\subsection{Density Estimator}

The great advantage of using the NFW profile is that the resulting
critical envelope depends exclusively on the bound mass of the
studied structure. In this way, the projected velocity envelope
can be scaled (changing either $r_\Delta$ or $M_\Delta$) until the
contained mass and volume satisfy some condition equivalent to the
critical density contrast, but in redshift space. For this
purpose, we will define an observable to characterize the
redshift-space density contrast inside the surface given by a
certain projected velocity envelope. The redshift-space ``density
estimator" is defined as
\begin{equation}
\label{dens-estim}
\hat{\Omega}_\mathrm{R}=\frac{\hat{M}_\mathrm{R}}{V_\mathrm{R}\rho_\mathrm{c}},
\end{equation}
where $\hat{M}_\mathrm{R}$ is the mass inside the projected
velocity envelope defined by the radius $R$ and $V_\mathrm{R}=4\pi
R^3/3$ is the real-space volume enclosed by the same radius. This
definition allows us to obtain density estimators for any
projected velocity envelope, including the critical envelope. In
practice we will need to define density estimators for only two
kinds of projected velocity envelopes: the critical envelope and
the ``core envelope", which is the projected velocity envelope
obtained using the NFW density profile for radii within
$r_\Delta$, necessary to estimate the NFW density profile in
redshift space. These density estimators can be computed for all
the structures in our sample, and then averaged to produce a
calibrated estimator that can let us infer the correct critical
envelope without need of real-space data.

The aim is to have a general template for the desired projected
velocity envelope, which can be associated to a specific density
estimator, specially calibrated for it. When fitting a template to
data from a redshift survey, it can be scaled until its density
estimator takes the desired value, yielding an estimation for the
radius in real space that characterises it.

To understand better the properties of the density estimator, we
calculated it using the ``true critical envelope'' (infered from
the true density obtained from the simulations) from all our 11
objects from 3 points of view. Results are shown in Table
\ref{DE}. The mean value for the density estimator is
$\hat{\bar{\Omega}}_\mathrm{cs} = 1.60$. This value is
significantly lower than the criterion $\Omega_\mathrm{cs} = 2.36$
for real space, implying that the selected volume is less
populated or, in other words, many more particles escaped the
theoretical prediction, also reflected in the greater value for
$C$. This leads to the conclusion that perturbations away from the
spherical collapse model have a stronger manifestation in
particles velocities than in their positions, decreasing the
reliability of predictions in redshift space.

\begin{table*}
  \caption{Density estimators for all projections. Rows: (1 - 3) Projections
in the $x$, $y$ and $z$ axes.  (4) Mean values for each object.
Columns: All 11 objects.}
  \centering
  \begin{small}
  \begin{tabular}{| l  c  c  c  c  c  c  c  c  c  c  c |}
  \hline
Object  & \#1 & \#2 & \#3 & \#4 & \#5 & \#6 & \#7 & \#8 & \#9 & \#10 & \#11\\
  \hline
 $x$ axis & 1.92  &  1.48  &  1.65  &  2.16  &  1.23  &  1.30  &  1.43  &  1.84  &  1.77  &  1.68  &  1.68\\
 $y$ axis & 1.54  &  1.81  &  1.63  &  1.62  &  1.34  &  1.20  &  1.37  &  1.50  &  1.77  &  1.81  &  1.67\\
 $z$ axis & 1.55  &  1.58  &  1.76  &  1.45  &  1.40  &  1.12  &  1.53  &  1.49  &  1.84  &  1.83  &  1.87\\
 Mean & 1.67  &  1.62  &  1.68  &  1.74  &  1.32  &  1.20  &  1.44  &  1.61  &  1.79  &  1.77  &  1.74\\
    \hline
  \end{tabular}
  \end{small}
  \label{DE}
\end{table*}

\subsection{Critical Envelope Recipes and Fitting Method}\label{recipes}

We would like to replace the true density profiles (unknown in
actual observed structures) by a simple parametrisation, based on
the NFW density profile. In order to test its accuracy, we compare
the previously defined performance indicators for the following
three projected velocity envelopes:

\begin{itemize}
\item True envelope: The projected velocity envelope is determined
directly from the true density profile.
\item NFW envelope: The
virialization radius $r_\Delta$ is determined using the
appropriate over-density criterion in real space, yielding an NFW
density profile, which is used to generate a projected velocity
envelope defined for shells within the critical radius according
to the NFW density profile.
\item Combined envelope: $r_\Delta$
and $r_\mathrm{cs}$ are found using the appropriate over-density
criteria in real space. We generate a projected velocity envelope
using the ellipsoid corresponding to the critical shell, adding
the inner part of the NFW envelope where it protrudes from the
ellipsoid.
\end{itemize}

Table \ref{Mean_stat_real} gives a comparison of the statistical
indicators for all the three critical envelopes in consideration.
We clearly see how the combined envelope shows almost the same
statistical properties as the true envelope. In contrast the NFW
envelope shows a greater amount of particles that escaped the
profile (Indicator $C$).

\begin{table*}
  \caption{Mean values and standard deviations for several performance
indicators for three critical envelopes obtained from real-space
measurements. Rows: (1) True: Critical envelope calculated using
the true density profile. (2) NFW: Uses the NFW density profile
according to $r_\Delta$ in real space. (3) Comb: Uses the NFW
profile for the innermost radii and the maximum ellipse given by
the critical radius. }
  \centering
  \begin{small}
  \begin{tabular}{| l  r  r  r  r  r  r  r  r |}
  \hline
  Envelope &  $\bar{A}$ & $\sigma_A$ & $\bar{B}$ & $\sigma_B$ & $\bar{C}$ & $\sigma_C$ & $\bar{D}$& $\sigma_D$\\
  \hline
  True   & 23.96 & 10.01 &  76.04 & 10.01 & 30.59 & 13.24 & 106.62 & 17.33\\
  NFW   & 19.72 &   7.35 &  80.28 &  7.35  &  37.17 & 21.67 & 117.42 & 24.23\\
  Comb & 23.88 &   9.98 &  76.12 &  9.98  &  30.80 & 14.09 & 106.93 & 17.55\\
    \hline
  \end{tabular}
  \end{small}
  \label{Mean_stat_real}
\end{table*}

Until now, the projected velocity envelopes have been determined
from information in real space that is not available in actual
observed structures. To fit the velocity envelopes directly in
redshift space, we will need to use density estimators in order to
scale the templates up to the desired size. Here we propose three
ways to do this.

The first way is to directly try to estimate the NFW density
profile for the structure, for which one needs to estimate the
virialization radius $r_\Delta$. In order to estimate $r_\Delta$
from the redshift space data, we defined another template based on
the NFW density profile:
\begin{itemize}
\item Core envelope:  equivalent to the NFW envelope, but now considering only shells inside the virialization radius.
\end{itemize}
Then we can construct a density estimator
$\hat{\Omega}_\mathrm{core}^\mathrm{NFW}$ that can be used to fit
the core envelope in redshift space. Once we have estimated
$r_\Delta$, we can use it for estimating
$r_\mathrm{cs}^\mathrm{NFW}$, defined as the radius where the
density contrast in real space
$\Omega_\mathrm{s}^\mathrm{NFW}=2.36$. We labeled this recipe as
``NFW-core".

The second way is to directly fit the NFW envelope, which is
defined up to $r_\mathrm{cs}^\mathrm{NFW}$, using a density estimator
previously calibrated for this template
($\hat{\Omega}_\mathrm{cs}^\mathrm{NFW}$). The fitted envelope
will directly estimate the critical radius $r_\mathrm{cs}^\mathrm{NFW}$. We
labeled this recipe as ``NFW-cs".

The last method is to use the combined envelope, for which we need
to estimate two independent parameters, $r_\Delta$ and
$r_\mathrm{cs}$. For the first we can fit the core envelope,
following the same procedure as in the NFW-core recipe. After
estimating $r_\Delta$, we can construct the combined envelope
leaving as a single free parameter the critical radius
$r_\mathrm{cs}$, which defines the outer ellipsoid. Finally we can
use another density estimator,
$\hat{\Omega}_\mathrm{cs}^\mathrm{comb}$, previously calibrated
for the combined envelope, to fit this last parameter. We labeled
this recipe as ``comb".

Having estimated the critical radius using one of these methods,
we use equation (\ref{real-space-crit}) to estimate the enclosed
mass. Then the true bound mass can be estimated using the
statistical relation between the mass inside the critical shell
and the bound mass, as explained in Paper I.

\subsection{Centre Determination}

A very important step in our method for identifying the structure
limits is to first have a good estimate for its centre, where to
anchor the projected velocity envelope. The intuitive way to do
this is to look for the region with the highest density of
particles (galaxies) in redshift space. In other words, we need to
find the centre which maximises the size of the projected velocity
envelope for a given density estimator. We performed this
maximisation using the core envelope, focusing on finding the
densest virialized centre for our structure. As density estimator
we used  $\hat{\Omega}_\mathrm{core}^\mathrm{NFW}$ calibrated for
the centre found in real space, using the procedure described in
the next subsection.

We found that the new centre appeared to be displaced from the
original centre on average a 13.5\% of $r_\Delta$ (representing
the virialized core), but the errors produced from using this new
centre were marginal compared to other systematic errors, being
around one order magnitude smaller.

\subsection{Density Estimator Calibration}\label{DEC}

In order to use the general projected velocity envelope to
estimate the critical radius of gravitationally bound structures,
we first need to obtain a density estimator specially calibrated
for that particular type of projected velocity envelope. The best
way of doing this is to use a statistically significant number of
simulated structures in the scale range we are interested in, and
calculate the corresponding density estimator using the actual
projected velocity envelope obtained from real-space measurements.

Here, we are going to use the mean of the density estimators found
in all 33 projections as our calibrated density estimator, and the
standard deviation will be used as the expected error. The density
estimators needed by our projected velocity envelope recipes are
$\hat{\Omega}_\mathrm{core}^\mathrm{NFW}$ for the core envelope,
$\hat{\Omega}_\mathrm{cs}^\mathrm{NFW}$ for the NFW envelope and
$\hat{\Omega}_\mathrm{cs}^\mathrm{comb}$ for the combined
envelope.

The mean density estimators, together with their standard
deviations are shown in Table \ref{Mean_rho_real}. For
completeness, and as a comparison point, we also included the mean
density estimator for the true envelope,
$\hat{\Omega}_\mathrm{cs}^\mathrm{true}$, which can also be found
for every single object and point of view in Table \ref{DE}.

\begin{table}
  \caption{Mean values and standard deviations for inner density estimators
for several projected velocity envelopes with parameters
determined in real space. See row definitions in the text.
The first three estimators correspond to envelopes defined
up to the critical radius, while the forth is defined up to the
much smaller virialization radius $\mathrm{r_\Delta}$.}
  \centering
  \begin{small}
  \begin{tabular}{| l l r  r  |}
  \hline
  Envelope &  Estimator & Mean & Std. Dev.\\
  \hline
  True     &   $\hat{\Omega}_\mathrm{cs}^\mathrm{true}$  &  1.59   &  0.25\\[3pt]
  NFW    &   $\hat{\Omega}_\mathrm{cs}^\mathrm{NFW}$  &     2.06   &  0.35\\[3pt]
  Comb  &   $\hat{\Omega}_\mathrm{cs}^\mathrm{comb}$  &     1.59   &  0.26\\[3pt]
  Core    &   $\hat{\Omega}_\mathrm{core}^\mathrm{NFW}$  &  109.64   & 14.56\\
    \hline
  \end{tabular}
  \end{small}
  \label{Mean_rho_real}
\end{table}

We observe that the True and Combined density estimators are very
similar, which tells us that the NFW profile accurately reproduces
the central part of the true velocity envelope. On the other hand,
the NFW envelope has a much greater density estimator, which is
the result of a smaller critical radius
($r_\mathrm{cs}^\mathrm{NFW}$). The density estimator associated
to the core envelope is actually very close to its equivalent
value in real space ($\Delta = 103.5$), which is good considering
that the structures' core are ruled by virialization, far from the
spherical collapse constraints we are assuming.

\subsection{Main Sources of Error}

\begin{figure*}
\centering
    \centering
    \includegraphics[width = 88mm, trim = 20 0 30 0]{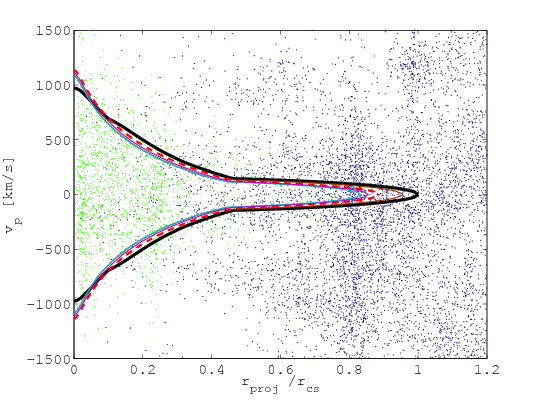}
    \includegraphics[width = 88mm, trim = 20 0 30 0]{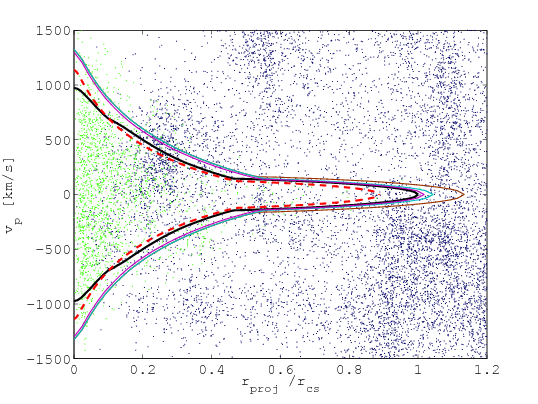}
    \caption{Projected velocities versus normalized projected radius for
Obj\#4. On the left is the $X$ axis view, while on the left is the
$Y$ axis view. Sampling 50\%.}
    \label{fig:perfiles4}
\end{figure*}

We can distinguish two kinds of error sources when first looking
at Table \ref{DE}: the first is associated to big variability in
the estimator depending on the axis being viewed, and the other is
manifested as an overall under or overestimation of the density
estimator for all views of the same object. For the first, the
axis dependence tells us that there is a strong influence of the
cluster's particular geometry or anisotropies outside of the
spherical distribution. A good example is Obj.\# 4, where we find
a big deviation in observations done along the $X$ axis, showing
an overestimation of the inner density with respect to
observations done along the other axes. After analyzing the
statistics, and plotting the radial velocities from all axes (see
Figure \ref{fig:perfiles4}), we conclude that this difference is
produced by an inflow of particles from an external structure
placed right on the line of sight, and whose virialized velocities
mix in redshift space with particles from the studied cluster. The
effect is an oversized parameter $A$ and an undersized parameter
$C$.

These observations appeared repeated in every object with big
standard deviation of the density estimator among the three axes,
but did not appear related to objects with highly non-spherical
cores or double cores.

\begin{figure*}
    \includegraphics[width = 88mm, trim = 20 0 30 0]{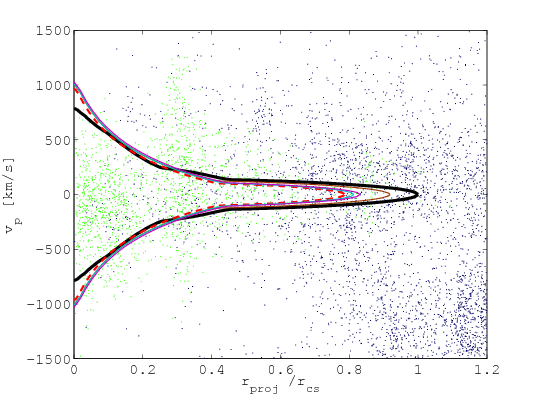}
    \includegraphics[width = 88mm, trim = 20 0 30 0]{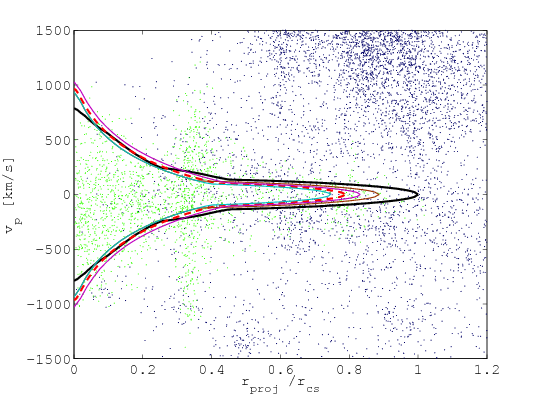}
    \caption{Projected velocities versus normalized projected radius for
Obj\#5. This object presents a double core that will eventually
merge into a single object. We observe that this feature does not
produce a significant effect when observing it from a different
point of view. Sampling 50\%.}
    \label{fig:perfiles5}
\end{figure*}

Regarding the second type of error, we conclude that it is related
to highly non-spherical cores or anisotropies inside the critical
radius. That is the case of Obj.\#5 shown in Figure
\ref{fig:perfiles5}, which has a double core\footnote{In the
analysis we used as centre the centre of the bigger core.}. We
observe that the double core does not produce a big deviation
between estimations from different axes, but produces a
significant underestimation of the inner density, which at the end
will cause that this kind of objects will be bigger than
predicted.

\section{Fitting Method Testing} \label{testing}

Before being able to apply the method to real observations, we
tested it in our simulations, in order to get statistical
information that can be used to interpret later results.

\subsection{Performance Statistics}

We fitted the three kinds of critical envelopes to our data in
redshift space using our previously calibrated density estimators.
Considering that we have only 11 objects in our study, we decided
to use the same data to do the fittings, but we rotated the
redshift projections in $45^\circ$ to at least use different
points of view at this stage. To fit the profiles, we just scaled
them until the density estimator condition was reached. As centre
for the profiles, we used our redshift-determined centres. We
realized three fittings:
\begin{itemize}
\item Combined method (comb): Uses the density estimator
$\hat{\Omega}_\mathrm{core}^\mathrm{NFW}$ to fit the core envelope
and calculate $r_\Delta$. Then fits an ellipsoid (that corresponds
to the critical shell) adjusting its radius until the combined
envelope produce the desired density estimator,
$\hat{\Omega}_\mathrm{cs}^\mathrm{comb}$, directly estimating
$r_\mathrm{cs}$. \item NFW-based method (NFW-core): Uses the
density estimator $\hat{\Omega}_\mathrm{core}^\mathrm{NFW}$ to fit
the core envelope and calculate $r_\Delta$. Then uses the NFW
density profile to calculate the critical radius
$r_\mathrm{cs}^\mathrm{NFW}$. \item NFW-based method 2 (NFW-cs):
Uses the density estimator $\hat{\Omega}_\mathrm{cs}^\mathrm{NFW}$
to fit the whole NFW envelope to the structure in redshift space,
directly estimating $r_\mathrm{cs}^\mathrm{NFW}$.
\end{itemize}

For the radius and mass estimations, we need to consider the
systematic biases discussed in subsection \ref{Fitting the Density
Profile} for the critical radius obtained from the NFW density
profile $r_\mathrm{cs}^\mathrm{NFW}$ and for the bound mass
estimated from the mass enclosed by the critical radius. In
particular we use that
$\hat{r}_\mathrm{cs}=r_\mathrm{cs}^\mathrm{NFW}/0.898$ and
$\hat{M}_\mathrm{bound}=0.717\hat{M}_\mathrm{cs}$, where
$\hat{M}_\mathrm{cs}$ is the mass derived from equation
(\ref{real-space-crit}).

There are two main statistics we are interested in:  The first
denotes mean systematic deviations of the estimated parameter
(radius or mass) from the true value\footnote{For the mass, the
true value is the bound mass in $a=100$.}. For this we define the
parameter $\alpha_\mathrm{i}$ as the mean of the ratio between the
estimated and the true value of the physical property $i$. The
existence of this kind of mean biases is mostly due to statistical
deviations, and should disappear as we increase the number of
structures studied for the whole calibration and testing of this
system.

In second place we are interested in quantifying errors due to
morphological characteristics that depart individual structures
from the mean. A way to do this is calculating the standard
deviation of measurements done from different angles, normalised
by the true value, which has the advantage of discounting the mean
bias produced by systematic errors. We labelled this as
$\sigma_{x,i}$, where $i$ denotes again the physical property we
are referring to. The mean parameter, averaged over all
structures, $\bar{\sigma}_x$, will give us the error we expect to
find in estimations done without taking into consideration the
particular morphological properties of the studied structure.

\begin{table}
  \caption{Systematic bias errors and mean standard deviation. Columns: (1):
Fitting method. (2), (4) and (6): Mean normalised bias
$\bar{\alpha}$ in the measured $r_\Delta$, $r_\mathrm{cs}$ and
$M$. (3), (5) and (7): standard deviation $\bar{\sigma}_x$, which
represents the deviation of estimations along different axes with
respect to the mean of the estimations for the same object.}
  \centering
  \begin{small}
  \begin{tabular}{| l  c  c  c  c  c  c |}
  \hline
Method & $\bar{\alpha}_{\Delta}$ & $\bar{\sigma}_{x,\Delta}$ & $\bar{\alpha}_{c}$  & $\bar{\sigma}_{x,c}$ & $\bar{\alpha}_M$ & $\bar{\sigma}_{x,M}$\\
  \hline
Comb   &   1.013  &   0.080   &    1.000  &   0.068   &    1.045   &   0.206\\
NFW1   &   1.013  &   0.080   &    1.007  &  0.083   &    1.086   &   0.253\\
NFW2   &   1.015  &   0.066   &    1.003  &   0.068    &    1.055   &   0.203\\
  \hline
  \end{tabular}
  \end{small}
  \label{Bias_error}
\end{table}

The mean results from this analysis can be found in Table
\ref{Bias_error}. There we can observe that the systematic biases
are all very small, reaching in the worst cases 1.5\% for
$r_\Delta$ estimations, 0.7\% for $r_\mathrm{cs}$ estimations and
8.6\% for mass estimations. This result should be taken with some
care, because we need to consider that there is an evident
correlation between the source and the test data for our system,
but this situation can be improved very easily by growing our
sample with new simulated structures.

For the standard deviation, we observe that errors due to
morphological anisotropies out of the spherical symmetry reach in
the worst cases 8.0\% for $r_\Delta$ estimations, 8.3\% for
$r_\mathrm{cs}$ estimations and 25\% for mass estimations.

\begin{figure*}
    \includegraphics[width = 58mm]{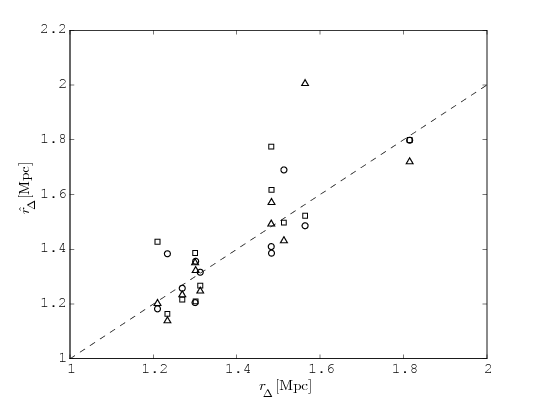}
    \includegraphics[width = 58mm]{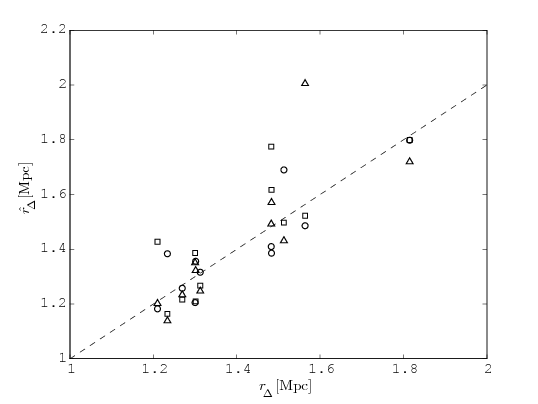}
    \includegraphics[width = 58mm]{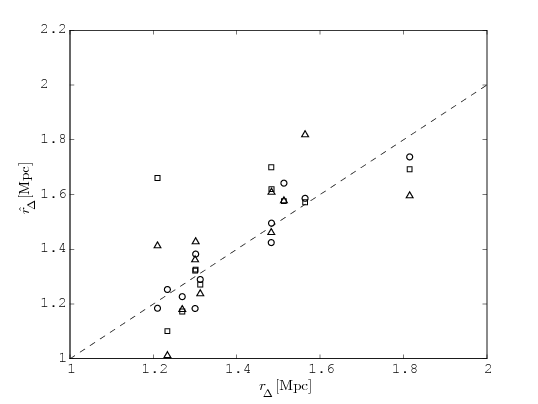}
    \caption{Estimated virial radii $r_\Delta$ versus measured virial radii for
all profile recipes and points of view. On the left, results from
the combined method. In the centre, results from NFW-core method.
On the right, results from NFW-cs method. Circles indicate
$X$-axis projections, triangles indicate $Y$-axis projections and
squares indicate $Z$-axis projections.}
    \label{Radii_Delta}

    \includegraphics[width = 58mm]{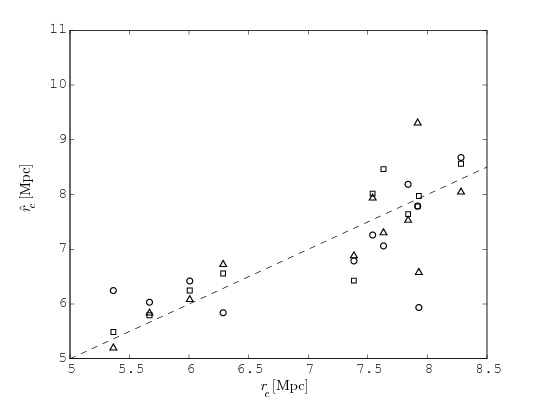}
    \includegraphics[width = 58mm]{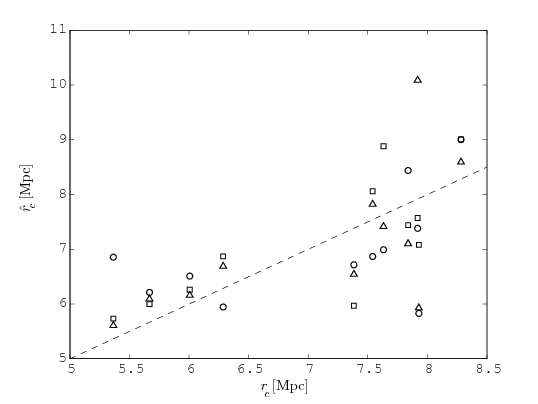}
    \includegraphics[width = 58mm]{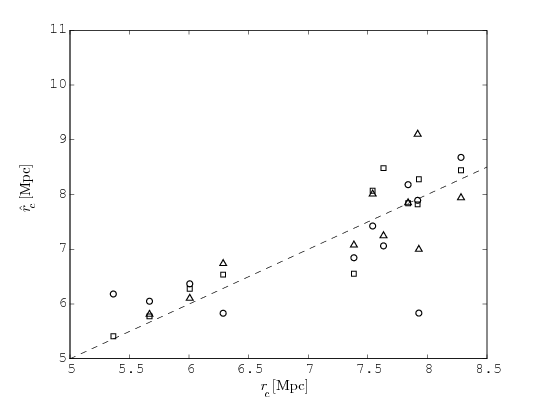}
    \caption{Estimated critical radii versus measured critical radii for all
profile recipes and points of view. On the left, results from the
combined method. In the centre, results from NFW-core method. On
the right, results from NFW-cs method. Circles indicate $X$-axis
projections, triangles indicate $Y$-axis projections and squares
indicate $Z$-axis projections. Results have been debiased
considering that $r_\mathrm{cs,NFW}=0.898r_\mathrm{cs}$}
    \label{Radii}

    \includegraphics[width = 58mm]{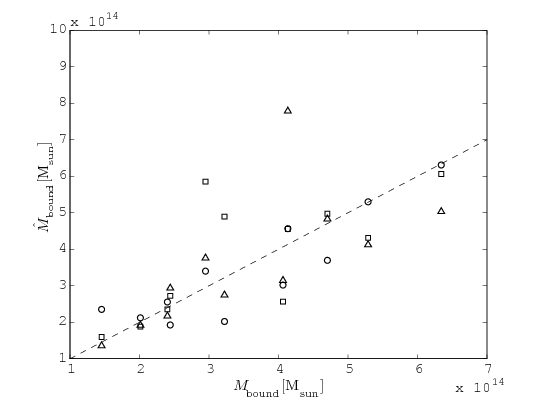}
    \includegraphics[width = 58mm]{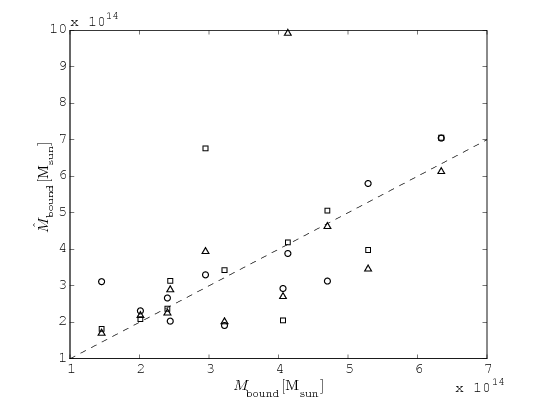}
    \includegraphics[width = 58mm]{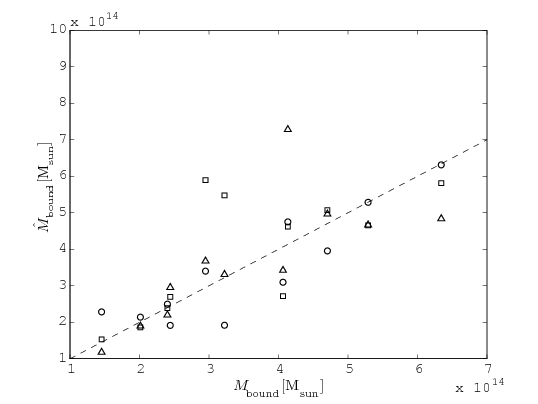}
    \caption{Estimated bound masses versus measured bound masses (in $a=1$). On
the left, results from the combined method. In the centre, results
from NFW-core method. On the right, results from NFW-cs method.
Circles indicate $X$-axis projections, triangles indicate $Y$-axis
projections and squares indicate $Z$-axis projections. Results
have been debiased considering that
$r_\mathrm{cs}^\mathrm{NFW}=0.898r_\mathrm{cs}$ for the NFW-based
methods and $M_\mathrm{bound}=0.717M_\mathrm{cs}$ for all.}
    \label{Masses}
\end{figure*}

In Figures \ref{Radii_Delta}, \ref{Radii} and \ref{Masses}, we
present the results for all methods. It is clear that the method
is axis-dependent, showing big deviations in structures with
strange morphologies. These deviations are in agreement with the
variability observed in the density estimators shown in Table
\ref{DE}, suggesting that much can be done identifying types of
structures and calibrating custom density estimators for them.

As a last conclusion, we observe that the best results are
obtained with the combined and NFW-cs methods, both of which have
in common that they are fitted to $r_\mathrm{cs}$ instead of
$r_\Delta$.

\subsection{Error Estimation}

The most direct way to estimate the expected error from estimating
the size and mass of a bound structure using this method is to
calculate the actual deviations from the true value over all
objects in our study. For the critical radius and the bound mass
we used the following equation:

\begin{equation}
Err_x = \sqrt{\frac{1}{N}\sum_i\frac{\left(\hat{x}_i-x_i\right)^2}{x_i^2}},
\label{error_r}
\end{equation}
where $x$ is the parameter on which we want to determine the error.

\begin{table}
  \caption{Expected errors in critical radius and bound mass estimations.
Columns: (1): Fitting method. (2): Percentage error in critical
radius estimation. (3): Percentage error in bound mass
estimation.}
  \centering
  \begin{small}
  \begin{tabular}{| l  r  r  |}
  \hline
Method & $Error_r$\% & $Error_M\%$\\
  \hline
Comb   &   8.5     &   31.3\\
NFW1   &   12.2  &   44.1\\
NFW2   &   8.1    &   30.9\\
  \hline
  \end{tabular}
  \end{small}
  \label{errores}
\end{table}

The expected errors calculated in this way are shown in Table
\ref{errores}. We emphasise that these errors are the ones we
should expect after a ``brute-force" application of the method,
having no consideration of the observable morphology of the
structure. Due to the statistical nature of the fitting method,
one would expect significant improvement in the expected errors
after a more careful analysis of individual cases, where different
``kinds" of structures, including structures with evident double
cores or highly contaminated by substructure, were considered.
This analysis is beyond the scope of this work, but can be
included in future works.

\section{Conclusions}\label{Conclusiones}

We have presented a new method to estimate the mass and radius of
gravitationally bound structures based solely on redshift
information present in redshift survey catalogs. The method is
based on the spherical collapse model (Paper I) and on the
important observation that the theoretical critical envelope
correctly follows the true critical envelope deep inside the
virialized centre of the simulated structures. We used the NFW
density profile \citep{NFW} to generate a template for the
critical envelope, which was calibrated using $N$-body
simulations.

The extension of the method to redshift space gave birth to a new
set of criteria that we called ``density estimators". These were
defined as the expected redshift-space density inside the
projected velocity envelope. Their calculation was completely
empirical, depending on statistical analysis of simulated
structures due to the considerable velocity dispersion found in
them. We observed that the main cause of velocity dispersion was
the gain of peculiar velocities due to local interaction between
in-falling objects and substructure. The study of substructure and
of the morphology of the studied structure can be of great help to
improve the error bars, since the density estimators show great
dependence on these properties. From this, we conclude that
numerical simulations can be used to emulate the particular
properties of the studied structures, producing ``custom-made"
methods to obtain the best results.

In contrast to the caustic method, our method forces a
limit to the radius of the structure, defined as the maximum
radius at which one should expect to find bound objects. The shape
of the velocity envelope though is less flexible than the caustics
shape, suggesting that a combination of both methods can give
better results, finding the envelope at lower radii using the
caustics where they are more defined, and using the fixed envelope
at higher radii to set a limit to the structure.

The more reliable methods from the study were the ``combined" and
the ``NFW-cs" methods. The procedure to apply the combined method
to a redshift data set is then the following:
\begin{itemize}
\item Calibrate data so that every element is accurately related
to a mass, so that the whole data set serves as a redshift-space
mass field. \item Identify the center of the structure by
maximising the number of particles inside the core envelope for a
given density estimator. \item Fit the NFW profile to the central
region (core) by adjusting $r_{\Delta}$ until the measured density
estimator is equal to the mean density estimator for the core,
given in Table \ref{Mean_rho_real}. \item Add the ellipsoid given
by $r_\mathrm{cs}$ to the critical envelope, and adjust
$r_\mathrm{cs}$ to the value $\hat{r}_\mathrm{cs}$, at which the
density estimator inside the total critical envelope is equal to
the mean density estimator for the combined method, given in Table
\ref{Mean_rho_real}. \item Given $\hat{r}_\mathrm{cs}$, calculate
the bound mass using equation \ref{real-space-crit}. \item The
fractional errors for $\hat{r}_\mathrm{cs}$  and
$\hat{M}_\mathrm{bound}$ are given in Table \ref{errores}.
\end{itemize}

In our next paper we will present the application of the method
presented here to estimate the size and bound mass of the Shapley
supercluster \citep{Proust}.

\section*{Acknowledgments} The authors thank Volker Springel for generously allowing the use
of GADGET2 before its public release and Kenneth J. Rines for
comments that improved the manuscript. R.~D. received support from
a CONICYT Doctoral Fellowship, R.~D. and A.~R. received support
from FONDECYT through Regular Projects 1020840 and 1060644, and A.
M. from the Comit\'e Mixto ESO-Chile. P.~A.~A. thanks LKBF and the
University of Groningen for supporting his visit to PUC. H.Q.
acknowledges partial support from the FONDAP Centro de
Astrof\'isica.

\label{lastpage}

\end{document}